\documentstyle[12pt]{article}

\newcommand{\be}{\begin{equation}}
\newcommand{\ee}{\end{equation}}
\newcommand{\bea}{\begin{eqnarray}}
\newcommand{\eea}{\end{eqnarray}}

\newcommand{\SMI}{
                  \vspace{-10cm}
                  \begin{flushright}
                  hep-th/9912242
                  \end{flushright}
                  \vspace{9.5cm}
                 }
\title{Metric Fluctuations in Brane Worlds}
\author{
M. G. Ivanov\thanks{mgi@mi.ras.ru, mgi@socrates.berkeley.edu}\\
{\small \it Department of Physics; University of California;
            94720-7300, Berkeley, CA , USA}\\
{\small \it Moscow Institute of Physics and Technology;}\\
{\small \it  Institutsky Per.9, Dolgoprudny, Moscow Reg., Russia}
\\ and \\
I. V. Volovich\thanks{volovich@mi.ras.ru}\\
{\small \it Steklov Mathematical Institute;
 Gubkin St.8, 117966, Moscow, Russia}\\
}
\date{January 19, 2000}
\begin{document}
\maketitle
\SMI
\begin{abstract}
Recently, a realization of the four-dimensional gravity on a brane
in five-dimensional spacetime has been discussed. Randall and Sundrum
have shown that the equation for the {\it longitudinal}
components of the metric fluctuations
admits a normalizable zero mode solution, which has been interpreted as
the localized gravity on the brane.
We find that equations, which include also
the {\it transverse} components  of the metric fluctuations,
have a zero-mode solution, which is not localized on the brane.
This indicates that  probably the effective theory
is unstable or, in other words, actually it is not four-dimensional
but five-dimensional.
Perhaps a modification of the proposal by using
matter fields can lead to the trapping of gravity to the brane.
\end{abstract}

\newpage

\section{Introduction}

Recently a realization of the four-dimensional gravity on a brane
in five-dimensional spacetime has been discussed \cite{rs1}-\cite{SSM}.
 Randall and Sundrum \cite{rs1,rs2}
have shown that the {\it longitudinal}
components of the metric fluctuations
satisfy to the quantum mechanical
equation with the potential, which includes
an {\it attractive} delta-function. As a result one has
a normalizable zero mode, which has been interpreted as
the localized gravity on the brane.

However in order to speak
about the localized gravity
one has to demonstrate that not only longitudinal
but also the transverse components of the metric
are confined to the brane. In this note we point out that  the
{\it transverse} components  of the metric fluctuations
satisfy to the  equation with the potential, which includes
the  {\it repulsive } delta-function.
There is a zero mode
solution, which is
not localized on the brane
and this indicates that there is  unstability
or, in other words, that  the effective
theory actually is not four-dimensional
but five-dimensional. Therefore it seems the original
proposal from \cite {rs1,rs2} does not lead to the
realization of the four-dimensional
gravity on the brane in five-dimensional spacetime.
Perhaps a modification of this proposal by using
matter fields can lead to the trapping of gravity to the brane.

\section{Metric perturbation}

The action has the form
\be
  I=\frac{1}{2}\int d^Dx\sqrt{|g|}
    \left(
      R+2\Lambda
    \right)+I_{Brane},
\ee

The RS solution is [1-3]
\bea\label{metr}
  ds^2&=&\frac{\eta_{MN}dx^M dx^N}{(k\sum_{i} |z^i|+1)^2},\\
\nonumber
 M,N&=&0,\dots,3+n,~~~~~
 i=1,\dots,n,~~~~~
 z^i=x^{i+3},~~~~~
 \mu,\nu=0,1,2,3,
\eea
where $D=n+4$ and $\eta^{MN}$ is a Minkowski metric
with the signature $(+,-,-,-,\dots)$.
 The metric (\ref{metr}) has singularities at
$z^i=0$ for any $i=1,\dots,n$,
which correspond to intersecting $n+2$-branes.
 Their intersection at $z^i=0$ for all $i=1,\dots,n$ is 3-brane, which
bears standard model fields and corresponds to observable
4-dimensional universe.
 The propagation of the longitudinal
 $(\mu,\nu)$-components of metric perturbation
in the background (\ref{metr}) was studied in the papers \cite{rs2,ah}.
 These perturbations are bounded in the vicinity of the 3-brane.
In this note we consider the perturbation of the transverse
($(i,j)$ and $(i,\mu)$) components.

We write the solution in the form [1-3]
\bea
  g_{MN}^0&=&H^{-2}\eta_{MN},\\
\label{H}
  H&=&k\sum_{i=1}^{n}|z^i|+1,
\eea
where
\be
  k^2=\frac{-2}{n(n+2)(n+3)}\Lambda,\label{kL}
\ee
and $\Lambda$ is the bulk cosmological constant.

Let us consider perturbations, which are parametrized
by $h_{MN}$ in the following way
\bea
  g_{MN}=H^{-2}\left(\eta_{MN}+h_{MN}\right)=H^{-2}\tilde g_{MN}
\eea
and fix the gauge
\be
  h_{MN}\eta^{MN}=0,~~
  \eta^{KM}\partial_K h_{MN}=0.\label{gr}
\ee
The Einstein tensor $G_{MN}=R_{MN}-\frac{1}{2}g_{MN}R$ is
\be
  G_{MN}=\tilde G_{MN}+
          (D-2)
          \left[
             \frac{\tilde\nabla_M\tilde\nabla_N H}{H}
            +\tilde g_{MN}\tilde g^{KL}
             \left\{
             -\frac{\tilde\nabla_K\tilde\nabla_L H}{H}
              +(D-1)\frac{\tilde\nabla_K H\tilde\nabla_L H}{2H^2}
             \right\}
          \right],
\ee
where all objects, which bear tilde are calculated by using
the metric
$\tilde g_{MN}$,
\be
  \tilde g ^{MN}=\eta^{MN}-h^{MN}=\eta^{MN}-\eta^{MK}h_{KL}\eta^{LN},
\ee
We compute all objects up to the first order in $h_{MN}$.
We obtain
\bea
 G_{MN}&=&\tilde G_{MN}+
          (D-2)
             \frac{\partial_M\partial_N H
                  -\tilde\Gamma^K_{MN}\partial_K H}{H}\\ \nonumber
           &+&(D-2)\tilde g_{MN}\tilde g^{KL}
             \left\{
             -\frac{\partial_K\partial_L H
                  -\tilde\Gamma^I_{KL}\partial_I H}{H}
              +(D-1)\frac{\partial_K H\partial_L H}{2H^2}
             \right\},
\eea
where
\be
 \tilde\Gamma^K_{MN}=\frac{\eta^{KL}}{2}
        \left(
          \partial_N h_{LM}+\partial_M h_{LN}-\partial_L h_{MN}
        \right).
\ee
Using the gauge conditions (\ref{gr})
\be
  \eta^{KL}\tilde\Gamma^I_{KL}=0
\ee
we get
\be
  G_{MN}=-\frac{1}{2}\Box h_{MN}+
          (D-2)
          \left[
             \frac{\partial_M\partial_N H
                  -\tilde\Gamma^K_{MN}\partial_K H}{H}
             +\tilde g_{MN}\tilde g^{KL}
             \left\{
              -\frac{\partial_K\partial_L H}{H}
              +(D-1)\frac{\partial_K H\partial_L H}{2H^2}
             \right\}
          \right],
\ee
where $\Box=\eta^{MN}\partial_M\partial_N$.
 Now by using Einstein equations
\be
  G_{MN}=\Lambda g_{MN}+T_{MN}^{branes}
\ee
we can identify $\partial\partial H$-terms with $T^{branes}$,
because they contribute only on the brane surfaces, and
using (\ref{H}), (\ref{kL}) we identify the $(\partial H)^2$-term
with $\Lambda g_{MN}$.
 Therefore one gets
\be
  -\frac{1}{2}\Box h_{MN}-(D-2)
      \frac{\partial_K H}{H}
      \tilde\Gamma^K_{MN}
  -(D-2)(D-1)\eta_{MN}h^{KL}\frac{\partial_K H\partial_L H}{2H^2}=0.
\ee

Finally we obtain the wave equation for metric perturbation in the form
\be\label{wv}
  \Box h_{MN}+(D-2)
      \frac{\partial_L H}{H}
      \eta^{KL}
      \left(
          \partial_N h_{KM}+\partial_M h_{KN}-\partial_K h_{MN}
      \right)
+(D-2)(D-1)\eta_{MN}h^{KL}\frac{\partial_K H\partial_L H}{H^2}=0.
\ee

\section{Propagation of metric perturbations}

Function $H$ does not depend on $x^\mu$.
It allows us to rewrite the equation (\ref{wv}) in the following form
\bea
\label{ij}
  \Box h_{ij}+(D-2)
      \frac{\partial_m H}{H}
      \eta^{mn}
      \left(
          \partial_j h_{ni}+\partial_i h_{nj}-\partial_n h_{ij}
      \right)
+(D-2)(D-1)\eta_{ij}h^{mn}\frac{\partial_m H\partial_n H}{H^2}=0,\\
\label{imu}
  \Box h_{i\mu}+(D-2)
      \frac{\partial_m H}{H}
      \eta^{mn}
      \left(
          \partial_\mu h_{ni}+\partial_i h_{n\mu}-\partial_n h_{i\mu}
      \right)=0,\\
\label{munu}
  \Box h_{\mu\nu}+(D-2)
      \frac{\partial_m H}{H}
      \eta^{mn}
      \left(
          \partial_\nu h_{n\mu}+\partial_\mu h_{n\nu}-\partial_n h_{\mu\nu}
      \right)
+(D-2)(D-1)\eta_{\mu\nu}h^{mn}\frac{\partial_m H\partial_n H}{H^2}=0.
\eea
To solve the system one can  solve equation (\ref{ij}) to find
$h_{ij}$, then substitute $h_{ij}$ into equation (\ref{imu}) to find
$h_{i\mu}$ and finally substitute $h_{i\mu}$ into
equation (\ref{munu}) to find
$h_{\mu\nu}$.

If $h_{i\mu}=0$, then
equation (\ref{munu}) coincides with the wave equation
derived in \cite{rs2,ah} for the longitudinal polarization of the
perturbation
\be
  \left(
      \Box-(D-2)
      \frac{\partial_m H}{H}
      \eta^{mn}\partial_n
  \right)
  h_{\mu\nu}=0.
\ee
This equation can be transformed into the wave equation with attractive
delta-function
\bea
  &&\left(
    \frac{\Box}{2}+V^{(-)}(z)
  \right)\hat h_{\mu\nu}=0,\\ \nonumber
  && V^{(-)}(z)=
  \frac{n(n+2)(n+4)k^2}{8H^2}-\frac{(n+2)k}{2H}\sum_j\delta(z^j),
\eea
where $\hat h=H^{-(n+2)/2} h$.
There is a bound state, which corresponds to the localized four-dimensional
gravity \cite{rs2,ah}.
The zero-mass state corresponds
to $\hat h=c H^{-(n+2)/2}e^{ipx}$,
$p_\mu p^\mu=0$, so
\be
  h_{\mu\nu}=c_{\mu\nu}e^{ipx},
\ee
where $c_{\mu\nu}$ is a constant polarization tensor.

\section{Non-longitudinal polarization in 5 dimensions}

In the simplest case of one extra dimension equations
(\ref{ij})-(\ref{munu})
acquire the following form
\bea
\label{555}
  \left(
      \Box-3\frac{\partial_5 H}{H}
      \partial_5
     -12\left(\frac{\partial_5 H}{H}\right)^2
  \right) h_{55}=0,\\
\label{55mu}
  \Box h_{5\mu}-3\frac{\partial_5 H}{H}
      \partial_\mu h_{55}=0,\\
\label{5munu}
  \left(
      \Box+3\frac{\partial_5 H}{H}
      \partial_5
  \right) h_{\mu\nu}-3\frac{\partial_5 H}{H}
   \left(\partial_\mu h_{5\nu}+\partial_\nu h_{5\mu}\right)
 +12\eta_{\mu\nu}h_{55}\left(\frac{\partial_5 H}{H}\right)=0.
\eea
The equation (\ref{555}) can be transformed into the
wave equation with the repulsive delta-function
\bea
  &&\left(
    \frac{\Box}{2}+V^{(+)}(z)
  \right)\hat h_{55}=0,\\ \nonumber
  && V^{(+)}(z)=-\frac{169k^2}{8H^2}+\frac{3k}{2}\delta(z),
\eea
where $\hat h=H^{3/2} h$ and we denote $x^M=(x^\mu,z)$.

Zero-mass state corresponds
to the solution $h\sim e^{ipx}$,
$p_\mu p^\mu=0$.
Let us set
\be\label{h55wave}
  h_{55}=0.
\ee
After the substitution of $h_{55}$ (\ref{h55wave})
into equation (\ref{55mu})
we have
\be\label{5muf}
  \Box h_{5\mu}=0.
\ee
Let us set
\be
 h_{5\mu}=c_\mu(z) e^{ipx},~~~~h_{\mu\nu}=\psi_{\mu\nu}(z)e^{ipx}.
\ee
Then, from (\ref{5muf}) and (\ref{5munu}) one gets
\bea
  c^{\prime\prime}_{\mu}=0,\\
  -\psi^{\prime\prime}_{\mu\nu}+3f
  \left(\psi'_{\mu\nu}-i(c_\mu p_\nu+c_\nu p_\mu)
  \right)=0,
\eea
where we denote $f(z)=\partial_5 H/H$.
As a simple explicite solution we take
\be
   \psi_{\mu\nu}=c_{\mu\nu}
   +i(c_\mu p_\nu+c_\nu p_\mu)z,~~~~
   c_{\mu\nu}=c_{\nu\mu}=const,~~~c_\mu=const.
\ee
To satisfy the gauge conditions (\ref{gr}) we have
also to set $c_\mu p^\mu=0$,
$c_{\mu\nu}\eta^{\mu\nu}=0$,
$c_{\mu\nu}p^\nu=0$.

Let us summarize our solution
\bea
 \nonumber
 h_{55}&=&0,\\ \nonumber
 h_{5\mu}&=&c_\mu e^{ipx},\\ \label{fin}
 h_{\mu\nu}&=&
            \left(c_{\mu\nu}+
             i(c_\mu p_\nu+c_\nu p_\mu)z
            \right) e^{ipx},\\
 \nonumber
 px&=&p_\mu x^\mu,~~~p_\mu p^\mu=0,~~~c_\mu p^\mu=0,~~~
        c_{\mu\nu}\eta^{\mu\nu}=0,~~~
        c_{\mu\nu}p^\nu=0,
\eea
where $c_\mu$, $c_{\mu\nu}$ and $p_\mu$ are constants.
We assume, of course, that one takes the real (or imaginary)
part of the above expressions.

To conclude, we obtain the zero mode
solution (\ref{fin})
of the equations for the metric
perturbation with non-longitudinal polarization.
One has a massless vector field $h_{5\mu}$
on the brane.
The perturbation (\ref{fin}) is not localized on
the brane because $h_{\mu\nu}$ depends linearly on $z$.
This indicates that  probably the effective theory
is unstable or, in other words, actually it is not
four-dimensional but five-dimensional.
Perhaps the coupling with matter fields \cite{GW}
could lead to the trapping of gravity to the brane.

We are grateful to I.Ya. Arefeva and B. Dragovich
for useful discussions.  I.V.V. is supported in part 
by INTAS grant 96-0698 and RFFI-99-01-00105.


\begin{thebibliography}{30}
\bibitem{rs1} L. Randall and R. Sundrum,
{\em A large mass hierarchy from small extra dimension},
hep-ph/9905221,
Phys.Rev.Lett. 83 (1999) 3370-3373
\bibitem{rs2} L. Randall and R. Sundrum,
{\em An alternative to compactification},
hep-th/9906064
\bibitem{ah} N. Arkani-Hamed, S. Dimopoulos, 
G. Dvali and N. Kaloper,
{\em Infinitely large new dimensions},
hep-th/9907209
\bibitem{ah0} N. Arkani-Hamed, S. Dimopoulos and G. Dvali,
{\em The hierarchy problem and new dimensions at a millimeter},
hep-ph/9803315,
Phys.Lett. B429 (1998) 263-272
\bibitem{GW} W. D. Goldberger and M. B. Wise,
{\em Modulus Stabilization with Bulk Fields},
hep-ph/9907447,
Phys.Rev.Lett. 83 (1999) 4922-4925
\bibitem{Gog} M. Gogberashvili,
{\em Gravitational Trapping for Extended Extra Dimension},
hep-ph/9908347
\bibitem{Hal} E. Halyo, 
{\em Localized Gravity on Branes in anti-de Sitter Spaces},
hep-th/9909127
\bibitem{CG}  A. Chamblin and G. W. Gibbons, 
{\em Supergravity on the Brane},
hep-th/9909130
\bibitem {Are} I. Ya. Aref'eva,
{\em High Energy Scattering in the Brane-World and Black Hole Production},
hep-th/9910269
\bibitem{R3} C. Csaki, M. Graesser, L. Randall and J. Terning,
{\em Cosmology of Brane Models with Radion Stabilization},
hep-ph/9911406
\bibitem{Rub} C. Charmousis, R. Gregory and V. A. Rubakov,
{\em Wave function of the radion in a brane world},
hep-th/9912160
\bibitem{visser} M. Visser,
{\em An exotic class of Kaliza-Klein models},
hep-th/9910093,
Phys.Lett. B159 (1985) 22
\bibitem{gs} J. Garriga and M. Sasaki,
{\em Brane-world creation and black holes},
hep-th/9912118
\bibitem{gt} J. Garriga and T. Tanaka,
{\em Gravity in the brane-world},
hep-th/9911055
\bibitem{greg} R. Gregory,
{\em Nonsingular global string compactifications},
hep-th/9911015
\bibitem{ck} A. G. Cohen and D. B. Kaplan,
{\em Solving the hierarchy problem with noncompact extra dimensions},
hep-th/9910132
\bibitem{vv} E. Verlinde and H. Verlinde,
{\em RG-flow, gravity and the cosmological constant},
hep-th/9912018
\bibitem{lr} J. Lykken and L. Randall,
{\em The shape of gravity},
hep-th/9908076
\bibitem{kaloper} N. Kaloper,
{\em Bent domain walls as braneworlds},
hep-th/9905210,
Phys.Rev. D60 (1999) 123506
\bibitem{SSM} M. Sasaki, T. Shiromizu and Kei-ichi Maeda,
{\em Gravity, Stability and Energy Conservation on 
the Randall-Sundrum Brane-World}, 
hep-th/9912233
\end{thebibliography}
\end{document}